\documentclass[a4paper]{jpconf}
\usepackage{graphicx}
 \usepackage{ifpdf,amssymb,amsmath,epsfig,bm,pifont}
 \usepackage{citesort}
\begin{document}
\title{On applications of {\sf Mathematica} Package {\sf ``FAPT''} in QCD}

\author{Viacheslav Khandramai}

\address{Gomel State Technical University, Gomel 246746, Belarus}

\ead{v.khandramai@gmail.com}

\begin{abstract}
We consider computational problems in the framework of nonpower Analityc Perturbation
Theory and Fractional Analytic Perturbation Theory that are the generalization of the standard QCD perturbation theory.
The singularity-free, finite couplings ${\cal
A}_{\nu}(Q^2),{\mathfrak A}_{\nu}(s)$ appear in these approaches
as analytic images of the standard QCD coupling powers
$\alpha_s^{\nu}(Q^2)$ in the Euclidean and Minkowski domains,
respectively.
We provide a package {\sf ``FAPT''} based on the system {\sf
Mathematica} for QCD calculations of the images ${\mathcal A}_{\nu}(Q^2)$, ${\mathfrak A}_{\nu}(s)$ up to N$^3$LO
of renormalization group evolution.
Application of these approaches to Bjorken sum rule ana\-ly\-sis
and $Q^2$-evolution of higher twist $\mu_4^{p-n}$ is considered.
\end{abstract}

\section{Introduction}
The QCD perturbation theory (PT) in the region of space-like
momentum transfer $Q^2 = -q^2 > 0$ is based on expansions in
a series in powers of the running coupling $\alpha_{\rm
s}(\mu^2=Q^2)$ which in the one-loop approximation is given by
$\alpha_{\rm s}^{(1)}(Q^2) = (4\pi/b_0)/L$ with $b_0$ being the
first coefficient of the QCD beta function, $L =
\ln(Q^2/\Lambda^2)$, and $\Lambda$ is the QCD
scale.
The one-loop solution $\alpha_{\rm s}^{(1)}(Q^2)$ has a pole
singularity at $L=0$ called the Landau pole. The $\ell$-loop
solution $\alpha_{\rm s}^{(\ell)}(Q^2)$ of the renormalization
group (RG) equation  has an $\ell$-root singularity of the type
$L^{-1/\ell}$ at $L=0$, which produces the pole as well in the
$\ell$-order term $d_\ell\,\alpha_{\rm s}^\ell(Q^2)$.
This prevents the application of perturbative QCD in the
low-momentum space-like regime, $Q^2\sim\Lambda^2$, with the
effect that hadronic quantities, calculated at the partonic level
in terms of a power-series expansion in $\alpha_{\rm s}(Q^2)$, are
not everywhere well defined.

In 1997, Shirkov and Solovtsov discovered couplings $\mathcal A_1(Q^2)$ free of
unphysical singularities in the
Euclidean region~\cite{SS}, and Milton and Solovtsov discovered couplings $\mathfrak A_1(s)$ in the Minkowski
region~\cite{MS97}. Due to the absence of singularities of these
couplings, it is suggested to use this systematic approach, called
Analytic Perturbation Theory (APT), for all $Q^2$ and $s$. The APT
yields a sensible description of hadronic quantities in QCD (see
reviews~\cite{SS99,Shi00,SS06}), though there are alternative
approaches to the singularity of effective charge in QCD --- in
particular, with respect to the deep infrared region
$Q^2<\Lambda^2$. One of the main advantages of the APT analysis is
much faster convergence of the APT nonpower series as
compared with the standard PT power series (see~\cite{Shi12}).
Recently, the analytic and numerical methods, necessary to perform
calculations in two- and three-loop approximations, were
developed~\cite{Mag99,KM03,Mag05}. The APT approach was applied to
calculate properties of a number of hadronic processes,
including the width of the inclusive $\tau$ lepton decay to
hadrons~\cite{MSSY00,MSS01,CvVa06,CKV09,Mag10}, the scheme and
renormalization-scale dependencies in the
Bjorken~\cite{MSS98,PST08} and Gross--Llewellyn
Smith~\cite{MSS98GLS} sum rules, the width of $\Upsilon$ meson
decay to hadrons~\cite{SZ05}, meson spectrum~\cite{BNPSS}, etc.

The generalization of APT for the fractional powers of an
effective charge was done in~\cite{BMS05,BMS06} and called the Fractional Analytic Perturbation Theory
(FAPT). The important
advantage of FAPT in this case is that the perturbative results
start to be less dependent on the factorization scale. This
reminds the results obtained with the APT and applied to the
analysis of the pion form factor in the $O(\alpha_{\rm s}^2)$
approximation, where the results also almost cease to depend on
the choice of the renormalization scheme and its scale (for a
detailed review see~\cite{AB08} and references therein). The
process of the Higgs boson decay into a $b\bar{b}$ pair of quarks
was studied within a FAPT-type framework in the Minkowski region
at the one-loop level in~\cite{BKM01} and within the FAPT at the
three-loop level in~\cite{BMS06}. The results on the resummation
of nonpower-series expansions of the Adler function of scalar
$D_S$ and a vector $D_V$ correlators within the FAPT were
presented in~\cite{BMS10}. The interplay between higher orders of
the perturbative QCD expansion and higher-twist contributions in
the analysis of recent Jefferson Lab data on the lowest moment of
the spin-dependent proton structure function, $\Gamma_1^{p}(Q^2)$,
was studied in~\cite{PSTSK09} using both the standard PT and APT/FAPT.
The FAPT technique was also applied to analyse the
structure function $F_2(x)$ behavior at small values of
$x$~\cite{CIKK09,KotKri10} and calculate binding energies and
masses of quarkonia~\cite{AC13}. All these successful applications
of APT/FAPT necessitate to have a reliable mathematical tool for
extending the scope of these approaches. In this paper, we present
the theoretical background which is necessary for the running of
${\mathcal A}_{\nu}[L]$ and ${\mathfrak A}_{\nu}[L]$ in
the framework of APT and its fractional generalization, FAPT, and
which is collected in the easy-to-use {\sf Mathematica} package
{\sf``FAPT''}~\cite{KB13}. This task has been partially realized
for APT as the \texttt{Maple} package \texttt{QCDMAPT} in~\cite{NeSi10} and as the
\texttt{Fortran} package \texttt{QCDMAPT\_F} in~\cite{NeSi11}. We have
organized {\sf``FAPT''} in the same manner as the well-known package {\sf``RunDec''}~\cite{CKS00}.
A few examples of APT and FAPT applications are given.

\section{Theoretical framework}
Let us start with the standard definitions used in {\sf ``FAPT''} for
standard PT calculations. The QCD running coupling,
$\alpha_\text{s}(\mu^2)=\alpha_\text{s}[L]$ with
$L=\ln[\mu^2/\Lambda^2]$, is defined through
\begin{eqnarray}
 \label{eq:beta}
  \frac{d \alpha_\text{s}[L]}{d L}
    &=&
    \beta\left(\alpha_\text{s}[L];n_f\right)
    \,\,=\,\,
   -\,\alpha_\text{s}[L]\,
     \sum_{k\ge0}b_k(n_f)\,
      \left(\frac{\alpha_\text{s}[L]}{4\pi}\right)^{k+1}\,,
\end{eqnarray}
where $n_f$ is the number of active flavours. The $\beta$-function
coefficients are given by (see~\cite{Beringer:1900zz})
\begin{eqnarray}
 b_0(n_f)
  &=& 11 - \frac{2}{3} n_f\,,
  \nonumber\\
 b_1(n_f)
  &=& 102 - \frac{38}{3} n_f\,,
  \nonumber \\
 b_2(n_f)
  &=& \frac{2857}{2}
    - \frac{5033}{18} n_f
    + \frac{325}{54} n_f^2\,,
  \nonumber \\
 b_3(n_f)
  &=& \frac{149753}{6}
    + 3564\,\zeta_3
    - \left[\frac{1078361}{162} + \frac{6508}{27}\,\zeta_3\right] n_f
  \nonumber\\
  & &
  \label{eq:beta.coef}
    + \left[\frac{50065}{162} + \frac{6472}{81}\,\zeta_3 \right] n_f^2
    + \frac{1093}{729}  n_f^3\,.
\end{eqnarray}
$\zeta$ is Riemann's zeta function.
We introduce the following notation:
\begin{eqnarray}
 \label{eq:ck.def}
  \beta_f\equiv \frac{b_0(n_f)}{4\pi}\,,\quad
  a(\mu^2;n_f)
   \equiv \beta_f\,\alpha_\text{s}(\mu^2;n_f)
   \quad\text{and}\quad
  c_k(n_f)
   \equiv \frac{b_k(n_f)}{b_0(n_f)^{k+1}}\,.
\end{eqnarray}
Then Eq.\,(\ref{eq:beta}) in the $l$-loop approximation can be rewritten as:
\begin{eqnarray}
 \label{eq:beta.norm}
  \frac{d a_{(\ell)}[L;n_f]}{d L}
    &=&
    -\,\left(a_{(\ell)}[L;n_f]\right)^{2}\,
     \left[1+\sum_{k\ge1}^{\ell}c_k(n_f)\,\left(a_{(\ell)}[L;n_f]\right)^{k}\right]\,.
\end{eqnarray}
In the one-loop ($\ell=1$) approximation ($c_k(n_f)=b_k(n_f)=0$ for all $k\geq1$)
we have the solution
\begin{eqnarray}
 \label{eq:a.1L}
  a_{(1)}[L]
   &=& \frac{1}{L}
\end{eqnarray}
with the Landau pole singularity at $L\to0$. In the two-loop
($\ell=2$) approximation ($c_k(n_f)=b_k(n_f)=0$ for all $k\geq2$)
the exact solution of Eq.~(\ref{eq:beta}) is also
known~\cite{GGK98}
\begin{eqnarray}
 \label{eq:a.2L}
  a_{(2)}[L;n_f]
   = \frac{-c_1^{-1}(n_f)}
          {1 + W_{-1}\left(z_W[L]\right)}
   \quad\text{with}\quad
  z_W[L]
  = -c_1^{-1}(n_f)\,e^{-1-L/c_1(n_f)}
     \,,
\end{eqnarray}
where $W_{-1}[z]$ is the appropriate branch of the Lambert function.

The three- ($c_k(n_f)=b_k(n_f)=0$ for all $k\geq3$) and
higher-loop solutions $a_{(\ell)}[L;n_f]$ can be expanded in
powers of the two-loop one, $a_{(2)}[L;n_f]$, as has been
suggested and investigated in~\cite{KM03,Mag05,Mag10}:
\begin{equation}
 \label{eq:a.lL_in_2L}
  a_{(\ell)}[L;n_f]
   = \sum_{n\geq1} C_{n}^{(\ell)}\,\left(a_{(2)}[L;n_f]\right)^n.
\end{equation}
The coefficients $C_{n}^{(\ell)}$ can be evaluated
recursively. As has been shown in~\cite{Mag05}, this expansion has
a finite radius of convergence, which appears to be sufficiently
large for all values of $n_{f}$ of practical interest. Note here
that this method of expressing the higher-$\ell$-loop coupling in
powers of the two-loop one is equivalent to the 't\,Hooft scheme,
where one puts by hand all coefficients of the $\beta$-function,
except $b_0$ and $b_1$, equal to zero and effectively takes into
account all higher coefficients $b_i$ by redefining perturbative
coefficients $d_i$ (see for more detail~\cite{GaKa11}).

The basic objects in the Analytic approach are the analytic
couplings in the Euclidian ${\mathcal A}_\nu^{(\ell)}[L;n_f]$ and
Minkowskian ${\mathfrak A}_\nu^{(\ell)}[L_s;n_f]$ domains
calculated with the spectral densities
{${\rho_{\nu}^{(\ell)}}(\sigma;n_f)$} which enter into the
K\"allen--Lehmann spectral representation:
 \begin{eqnarray}
  \label{eq:A_1}
   {\mathcal A}_\nu^{(\ell)}[L;n_f]
   \!&\!=\!&\! \int_0^{\infty}\!\frac{\rho_\nu^{(\ell)}(\sigma;n_f)}{\sigma+Q^2}\,
                d\sigma
        =      \int_{-\infty}^{\infty}\!\frac{\rho_\nu^{(\ell)}[L_\sigma;n_f]}{1+\exp(L-L_\sigma)}\,
                dL_\sigma\,,\\
  \label{eq:U_1}
   {\mathfrak A}_\nu^{(\ell)}[L_s;n_f]
   \!&\!=\!&\! \int_s^{\infty}\!\frac{\rho_\nu^{(\ell)}(\sigma;n_f)}{\sigma}\,
                d\sigma
            = \int_{L_s}^{\infty}\!\rho_\nu^{(\ell)}[L_\sigma;n_f]\,
               dL_\sigma\,.
\end{eqnarray}

It is convenient to use the following representation for spectral
functions:
\begin{eqnarray}
 \rho_{\nu}^{(\ell)}[L;n_f]
  \equiv \frac{1}{\pi}\,
     \textbf{Im}{}
      \left(\alpha_\text{s}^{(\ell)}\left[L-i\pi;n_f\right]
      \right)^{\nu}
  = \frac{\sin[\nu\,\varphi_{(\ell)}[L;n_f]]}
         {\pi\,(\beta_f\,R_{(\ell)}[L;n_f])^{\nu}}\,.
 \label{eq:SpDen.lL.nu}
\end{eqnarray}

In the one-loop approximation the corresponding functions have the simplest form
\begin{eqnarray}
 \varphi_{(1)}[L]
  = \arccos\left(\frac{L}{\sqrt{L^2+\pi^2}}\right)\,,~~
 R_{(1)}[L]
  = \sqrt{L^2+\pi^2}\,,
 \label{eq:SpDen.1L.n}
\end{eqnarray}
whereas at the two-loop order they have a more complicated form
\begin{eqnarray}
 \label{eq:Lamb.R_(2)}
  R_{(2)}[L;n_f]
   &=& c_1(n_f)\,\Big|1+W_{1}\left(z_W[L-i\pi;n_f]\right)\Big|\,,\\
 \label{eq:Lamb.phi_(2)}
  \varphi_{(2)}[L;n_f]
   &=& \arccos
       \left[
        \textbf{Re}\left(
                   \frac{-R_{(2)}[L;n_f]}
                        {1+W_{1}\left(z_W[L-i\pi;n_f]\right)}
                   \right)
      \right]
\end{eqnarray}
with $W_{1}[z]$ being the appropriate branch of Lambert function.

In the three- ($\ell=3$) and four-loop ($\ell=4$) approximation we
use Eq.\,(\ref{eq:a.lL_in_2L}) and then obtain
\begin{eqnarray}
 \label{eq:Lamb.R_(3)}
  R_{(\ell)}[L]
   &=& \left|\frac{e^{i\,\varphi_{(2)}[L]}}{R_{(2)}[L]}
             + \sum_{k\geq3}
                C_{k}^{(\ell)}\,\frac{e^{i\,k\,\varphi_{(2)}[L]}}{R_{(2)}^k[L]}
       \right|^{-1}\,,~~~\\
 \label{eq:Lamb.phi_(3)}
  \varphi_{(\ell)}[L]
   &=& \arccos
      \left[\frac{R_{(\ell)}[L]\cos\left(\varphi_{(2)}[L]\right)}
                 {R_{(2)}[L]}
         + \sum_{k\geq3}
            C_{k}^{(\ell)}\,
             \frac{R_{(\ell)}[L]\cos\left(k\,\varphi_{(2)}[L]\right)}
                  {R_{(2)}^k[L]}
      \right]\,.
\end{eqnarray}
Here we do not show explicitly the $n_f$ dependence of the
corresponding quantities --- it goes inside through
$R_{(2)}[L]=R_{(2)}[L;n_f]$,
$\varphi_{(2)}[L]=\varphi_{(2)}[L;n_f]$,
$C_{k}^{(3)}=C_{k}^{(3)}[n_f]$, $C_{k}^{(4)}=C_{k}^{(4)}[n_f]$,
$c_{k}=c_{k}(n_f)$.

The package {\sf ``FAPT''} performs the calculations of the basic
required objects: $\left({\alpha}_{\rm s}^{(\ell)}[L, n_f]\right)^{\nu}$~in
Eqs.~(\ref{eq:a.1L}), (\ref{eq:a.2L}) and (\ref{eq:a.lL_in_2L}),
${\mathcal A}_\nu^{(\ell)}[L, n_f]$ in Eq.~(\ref{eq:A_1}) and
${\mathfrak A}_\nu^{(\ell)}[L,n_f]$ in Eq.~(\ref{eq:U_1}) up
to the N$^3$LO approximation ($\ell=4$) with a fixed number of active
flavours $n_f$ and the global one with taking into account all
heavy-quark thresholds (for more details and
description of procedures see~\cite{KB13}). As an example, we present here the
following {\sf Mathematica} realizations for analytic coupling
${\mathcal A}_\nu^{(\ell)}[L, n_f]$ and ${\mathfrak
A}_\nu^{(\ell)}[L, n_f]$:
\begin{itemize}
\item
\texttt{AcalBar}$\ell$\texttt{[L,Nf,Nu]} computes the $\ell$-loop
$n_f$-fixed analytic coupling ${\mathcal A}_\nu^{(\ell)}[L,
n_f]$ in the Euclidean domain, where the logarithmic argument
\texttt{L}=$\ln[Q^2/\Lambda^2]$, the number of active flavors
\texttt{Nf}=$n_f$, and the power index \texttt{Nu}=$\nu$;
\item
\texttt{UcalBar}$\ell$\texttt{[L,Nf,Nu]} computes the $\ell$-loop
$n_f$-fixed analytic coupling ${\mathfrak A}_\nu^{(\ell)}[L,
n_f]$ in the Minkowski domain, where the logarithmic argument
\texttt{L}=$\ln[s/\Lambda^2]$, the number of active flavors
\texttt{Nf}=$n_f$, and the power index \texttt{Nu}=$\nu$.
\end{itemize}

\section{APT and FAPT applications}
\begin{figure}[!h]\leavevmode\begin{center}
         \begin{minipage}[b]{0.49\textwidth}
                \phantom{}\hspace{-0.5cm}%
\centering\includegraphics[width=1.01\textwidth]{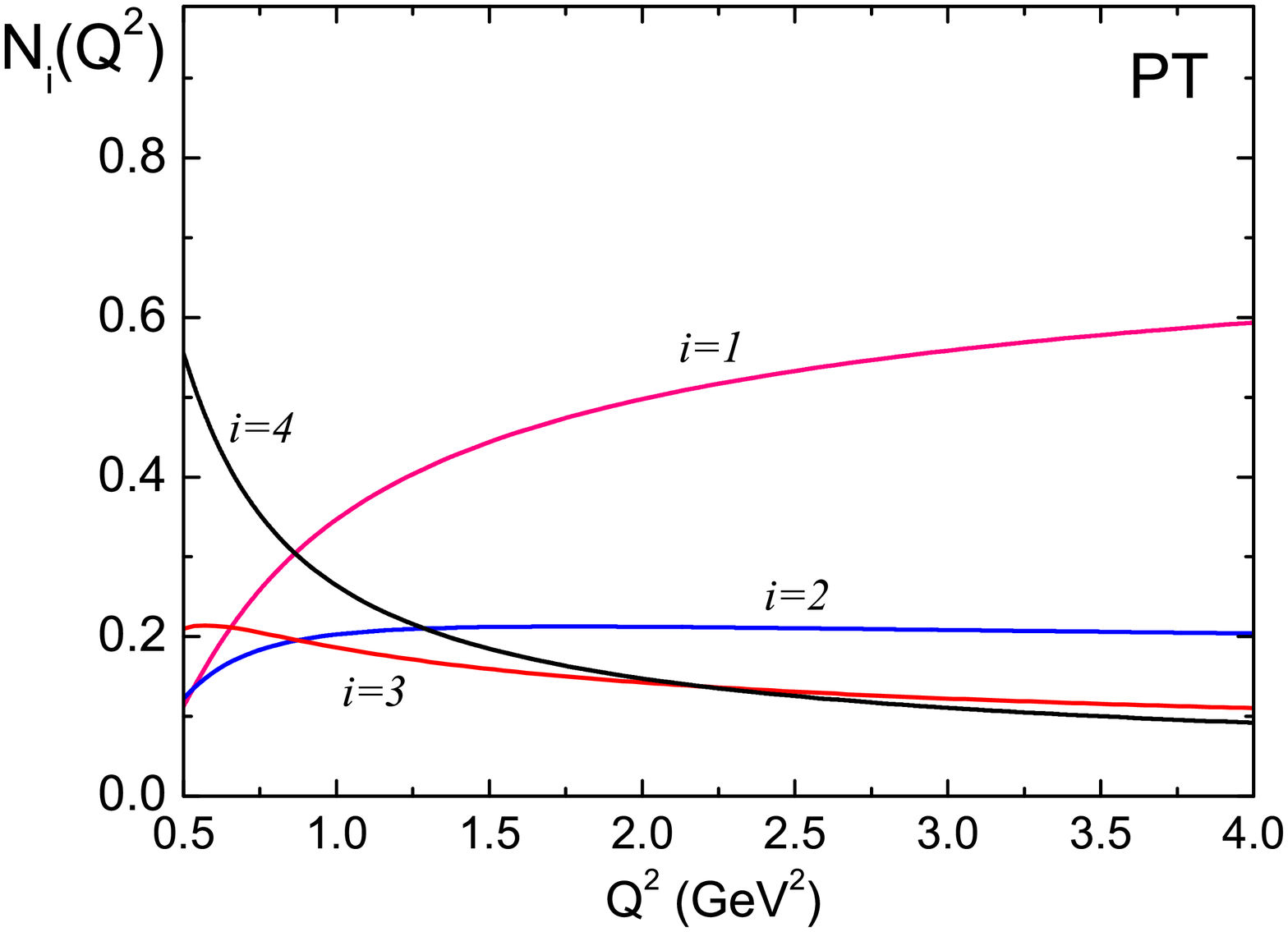}
         \end{minipage}%
     \begin{minipage}[b]{0.49\textwidth}
     \vspace*{-0.0cm}
\centering\includegraphics[width=1.01\textwidth]{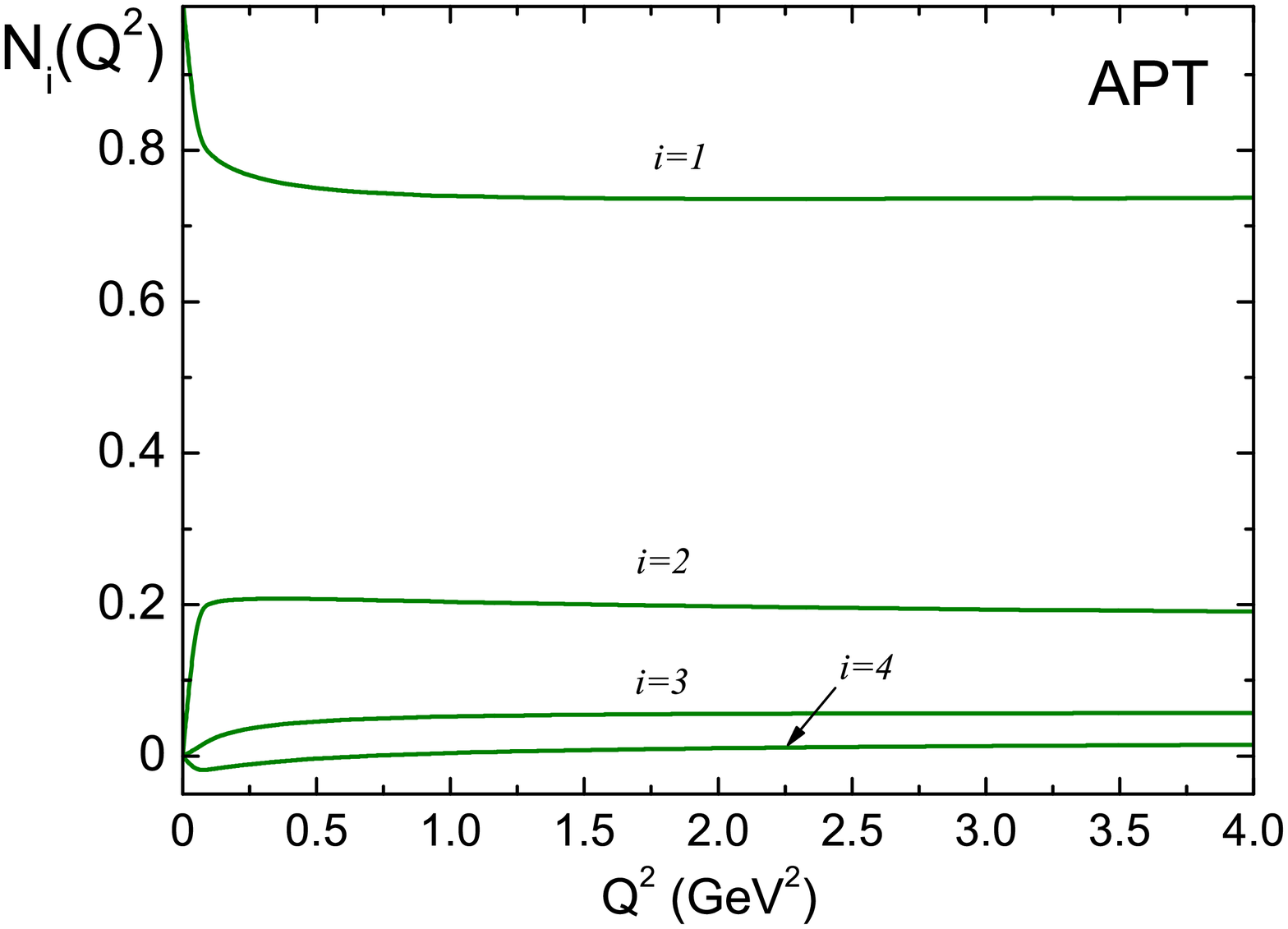} 
    \end{minipage}
             \begin{minipage}[t]{0.45\textwidth}
 \caption{\footnotesize The $Q^2$-dependence of the relative contributions at the four-loop
level in the PT approach.} \label{fig:s_PT}
        \end{minipage}%
\phantom{}\hspace{0.5cm}%
     \begin{minipage}[t]{0.45\textwidth}
 \caption{\footnotesize The $Q^2$-dependence of the relative contributions of the
perturbative expansion terms in Eq.~(\ref{Delta_APT-Bj}) in the APT
approach.} \label{fig:s_APT}
    \end{minipage}      \end{center}
    \end{figure}

As an example of the APT application, we present the Bjorken sum rule
(BSR) analysis (see for more details~\cite{KPSST12}). The BSR
claims that the difference between the proton and neutron structure
functions integrated over all possible values
\begin{equation}\label{Bj}
\Gamma^{p-n}_1(Q^2) \,=\, \int_0^{1}\,\ \left[g_1^p(x,Q^2) -
g_1^n(x,Q^2) \right] dx \, ,
\end{equation}
of the Bjorken variable $x$ in the limit of large momentum
squared of the exchanged virtual photon at $Q^2 \to \infty$ is
equal to $g_A/6$, where the nucleon axial charge $g_A=1.2701\pm0.0025$~\cite{Beringer:1900zz}.
Commonly, one represents the Bjorken integral in Eq. (\ref{Bj}) as
a sum of perturbative and higher twist contributions
\begin{eqnarray} \label{PT-Bj-HT}
\Gamma^{p-n}_1(Q^2)=\frac{g_A}{6}\biggl[1-\Delta_{\rm Bj}(Q^2)
\biggr]+\sum_{i=2}^{\infty}\frac{\mu^{p-n}_{2i}}{Q^{2i-2}} \,.
\end{eqnarray}
The perturbative QCD correction $\Delta_{\rm Bj}(Q^2)$ has a form
of the power series in the QCD running coupling $\alpha_{\rm
s}(Q^2)$. At the up-to-date four-loop level in the massless case
in the modified minimal subtraction ($\overline{\rm{MS}}$) scheme,
for three active flavors, $n_f=3$, it looks like~\cite{BCK}
\begin{equation}\label{Delta_PT-Bj}
 \Delta_{\rm Bj}^{\rm PT}(Q^2) = 0.3183\,\alpha_{\rm s}(Q^2)\,+\,0.3631\,\alpha_{\rm s}^2(Q^2)+0.6520\,\alpha_{\rm s}^3(Q^2)+\,1.804\,\alpha_{\rm s}^4(Q^2).
\end{equation}
The perturbative representation (\ref{Delta_PT-Bj}) violates
analytic properties due to the unphysical singularities of
$\alpha_{\rm s}(Q^2)$. To resolve the issue, we apply APT. In
particular, the four-loop APT expansion for the perturbative part
$\Delta_{\rm Bj}^{\rm PT}(Q^2)$ is given by the formal replacement
\begin{equation}
\label{Delta_APT-Bj}
 \Delta_{\rm Bj}^{\rm PT}(Q^2) = \sum_{k \leq 4}\,\, c_k \,\alpha_{\rm s}^k(Q^2) \quad \Rightarrow \quad \Delta_{\rm Bj}^{\rm APT}(Q^2)= \sum_{k \leq 4}\,\, c_k \,{\mathcal A}_k(Q^2)\,.
\end{equation}
Clearly, at low $Q^2$ a value of $\alpha_{\rm s}$ is quite
large, questioning the convergence of perturbative QCD
series~(\ref{Delta_PT-Bj}). The qualitative resemblance of the
coefficients pattern to the factorial growth did not escape our
attention although more definite statements, if possible,
would require much more efforts. This observation allows one to
estimate the value of $\alpha_{\rm s} \sim 1/3$ providing a
similar magnitude of three- and four- loop contributions to the
BSR. To test that, we present in Figs.~\ref{fig:s_PT}
and~\ref{fig:s_APT} the relative contributions of separate
$i$-terms in the four-loop expansion in Eq.~(\ref{Delta_PT-Bj})
for the PT case and in Eq.~(\ref{Delta_APT-Bj}) for APT.
As it is seen from Fig.~\ref{fig:s_PT}, in the region
$Q^2<1\,\text{GeV}^2$ the dominant contribution to the pQCD
correction $\Delta_{\rm Bj}(Q^2)$ comes from the four-loop term
$\sim\alpha_{\rm s}^4$. Moreover, its relative contribution
increases with decreasing $Q^2$. In the region
$Q^2>2\,\text{GeV}^2$ the situation changes -- the major
contribution comes from one- and two-loop orders there. Analogous
curves for the APT series given by Eq.~(\ref{Delta_APT-Bj}) are
presented in Fig.~\ref{fig:s_APT}.

Figures~\ref{fig:s_PT} and \ref{fig:s_APT} demonstrate the essential
difference between the PT and APT cases, namely, the APT expansion
obeys much better convergence than the PT one. In the APT case, the
higher order contributions are stable at all $Q^2$ values, and the
one-loop contribution gives about 70 \%, two-loop -- 20 \%,
three-loop -- not exceeds 5\%, and four-loop -- up to 1 \%.

One can see that the four-loop PT correction becomes equal to the
three-loop one at $Q^2=2\,\text{GeV}^2$ and noticeably overestimates it
(note that the slopes of these contributions are quite close in the
relatively wide $Q^2$ region) for $Q^2 \sim 1\,\text{GeV}^2$ which may be
considered as an extra argument supporting an asymptotic character
of the PT series in this region. In the APT case, the contribution of the higher loop corrections is
not so large as in the PT one. The four-loop order in APT can be
important, in principle, if the theoretical accuracy to better than
1 \% will be required.

\begin{figure}[!h]\leavevmode\begin{center}
         \begin{minipage}[b]{0.49\textwidth}
                \phantom{}\hspace{-0.5cm}%
\centering\includegraphics[width=0.98\textwidth]{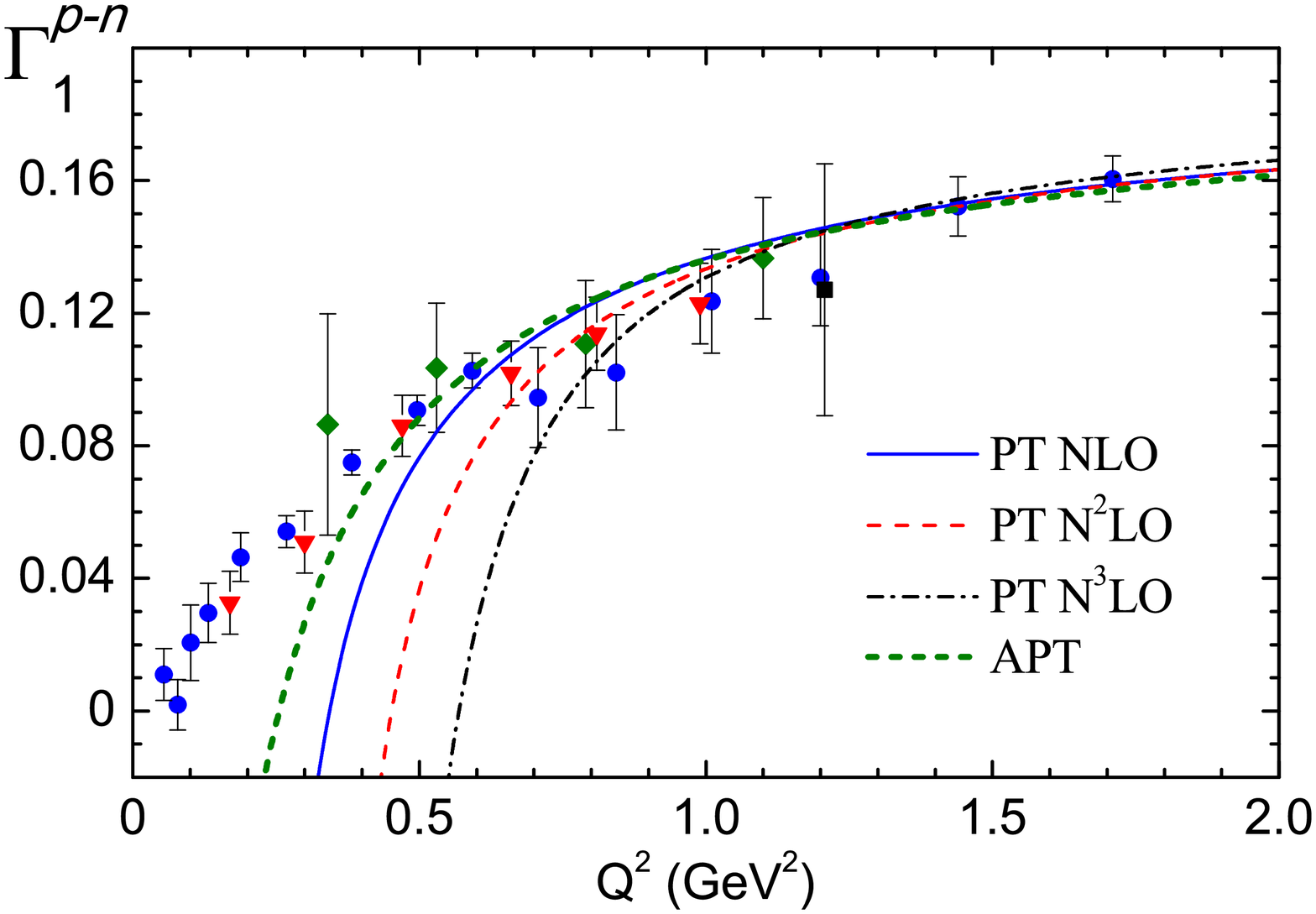}
         \end{minipage}%
     \begin{minipage}[b]{0.49\textwidth}
     \vspace*{-0.0cm}
\centering\includegraphics[width=0.98\textwidth]{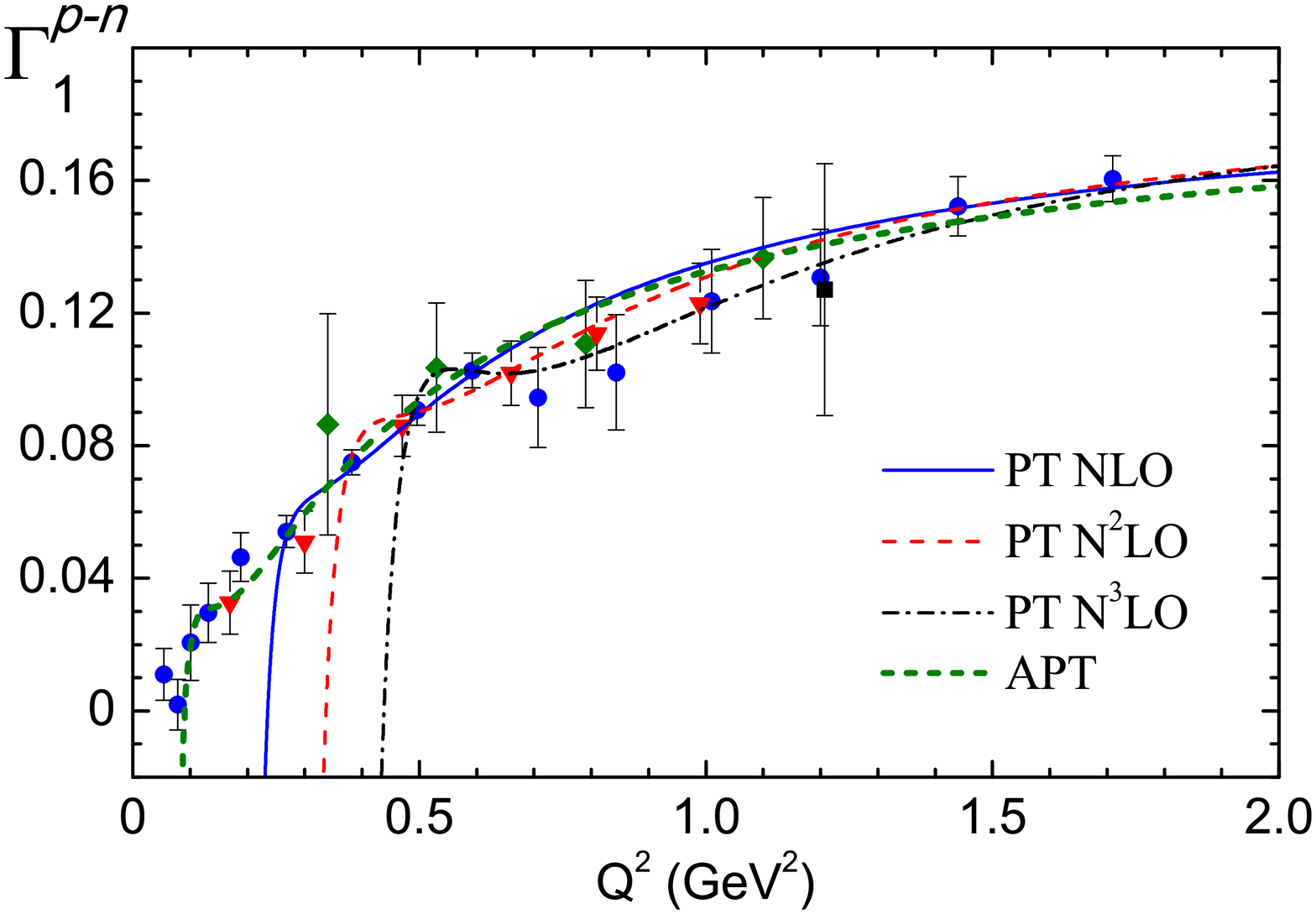} 
    \end{minipage}
             \begin{minipage}[t]{0.45\textwidth}
 \caption{\footnotesize The one-parametric $\mu^{p-n}_4$-fits of the BSR JLab data in
various (NLO, N$^2$LO, N$^3$LO) orders of the PT and the all-order
APT expansions.} \label{fig:fit1}
        \end{minipage}%
\phantom{}\hspace{0.5cm}%
     \begin{minipage}[t]{0.45\textwidth}
 \caption{\footnotesize The three-parametric $\mu^{p-n}_{4,6,8}$-fits of the BSR JLab data in
various (NLO, N$^2$LO, N$^3$LO) orders of the PT and the all-order
APT expansions.} \label{fig:fit2}
    \end{minipage}      \end{center}
    \end{figure}

Now we briefly discuss how the APT applications affect the
values of the higher-twist coefficients $\mu_{2i}^{p-n}$ in
Eq.~(\ref{PT-Bj-HT}) extracted from Jlab data. Previously, a
detailed higher-twist analysis of the four-loop expansions in
powers of $\alpha_{\rm s}$ was performed in~\cite{KPSST12}. In
Figs.~\ref{fig:fit1} and~\ref{fig:fit2} we present the results of
1- and 3-parametric fits in various orders of the PT and APT. The
corresponding fit results for higher twist terms
$\mu_{2i}^{p-n}$, extracted in different orders of the PT and APT, are
given in Table~\ref{tab:Bjtot_PT} (all numerical results are
normalized to the corresponding powers of the nucleon mass $M$).
    \begin{table}[h]
\caption{\label{tab:Bjtot_PT} Results of higher twist extraction from the JLab data on
BSR in various (NLO, N$^2$LO, N$^3$LO) orders of the PT and all orders
of APT.}
\begin{center}
\begin{tabular}{lcccc}
\br
Method& $Q_{min}^2,\,$ & $\mu^{p-n}_4/M^2$ &
$\mu^{p-n}_6/M^4$ & $\mu^{p-n}_8/M^6$  \\
 \br
\multicolumn{5}{c}{The best $\mu^{p-n}_4$-fit results}\\
\mr
PT NLO          &  ~$0.5$   & $-0.028(5)$  & $ - $        &  $ - $       \\
PT N$^2$LO      &  ~$0.66$  & $-0.014(7)$  &   $ -  $     &   $ - $      \\
PT  N$^3$LO     &  ~$0.71$  & $~~0.006(9)$ & $ -$      & $-$  \\
APT             & $~0.47$   & $-0.050(4)$  & $-$ & $-$  \\
\mr
\multicolumn{5}{c}{The best $\mu^{p-n}_{4,6,8}$-fit results}  \\
\mr
PT  NLO         & $~0.27$  & $-0.03(1)$   & $-0.01(1)$   & $0.008(4)$   \\
PT N$^2$LO      &  ~$0.34$ & $~~0.01(2)$  & $-0.06(4) $  & $0.04(2)~$   \\
PT N$^3$LO      &  ~$0.47$ & $~~0.05(4)$  &  $-0.2(1)~$  & $0.12(6)~$    \\
APT             & $~0.08$  & $-0.061(4)$  & $0.009(1)$   & $-0.0004(1)$  \\
\br
\end{tabular}
\end{center}
\end{table}
From these figures and Table~\ref{tab:Bjtot_PT} one can see that
APT allows one to move down up to $Q^2 \sim 0.1\,\text{GeV}^2$ in
description of the experimental data~\cite{KPSST12}. At the same
time, in the framework of the standard PT the lower border shifts
up to higher $Q^2$ scales when increasing the order of the PT
expansion. This is caused by extra unphysical singularities in the
higher-loop strong coupling. It should be noted that the magnitude of
$\mu^{p-n}_4/M^2$ decreases with an order of the PT and becomes
compatible to zero at the four-loop level. It is interesting to mention that a
similar decreasing effect has been found in the analysis of the
experimental data for the neutrino-nucleon DIS structure function
$xF_3$~\cite{KatParSid} and for the charged lepton-nucleon DIS
structure function $F_2$~\cite{Blum}.

Consider the application of the FAPT approach by the example of the
RG-evolution of the non-singlet higher-twist $\mu^{p-n}_4(Q^2)$ in
Eq.~(\ref{PT-Bj-HT}). The evolution of the higher-twist terms
$\mu^{p-n}_{6,8,\,...}$ is still unknown. The RG-evolution of
$\mu^{p-n}_4(Q^2)$ in the standard PT reads
\begin{eqnarray}\label{Eq:mu-evol}
\mu_{4,PT}^{p-n}(Q^2)&=&\mu_{4,PT}^{p-n}(Q_0^2)\left[\frac{\alpha_{\rm s}(Q^2)}{\alpha_{\rm s}(Q_0^2)}\right]^{\nu}\,,\quad\nu=\gamma_0/\left(8\pi\beta_0\right)\,,\quad \gamma_0=\frac{16}{3} C_F\,,\quad C_F=\frac{4}{3}.
\end{eqnarray}
In the framework of FAPT the corresponding expression reads as follows:
\begin{eqnarray}
\mu_{4,APT}^{p-n}(Q^2)= \mu_{4,APT}^{p-n}(Q_0^2)\,
\frac{{\mathcal A}_{\nu}^{(1)}(Q^2)}{{\mathcal A}_{\nu}^{(1)}(Q_0^2)}.
\label{APT-evol-mu}
\end{eqnarray}
We present in Table~\ref{tab:BjEv} the best fits for
$\mu^{p-n}_4(Q_0^2)$ taking into account the corresponding
RG-evolution with $Q^2_0=1\,\text{GeV}^2$ as a normalization
point and without the RG-evolution.
\begin{table}[h]
\caption{\label{tab:BjEv} Results of higher twist extraction from the JLab data on
BSR with inclusion and without inclusion of the RG-evolution of $\mu_4^{p-n}(Q^2)$ normalized at $Q^2_0=1\,\text{GeV}^2$.}
\begin{center}
\begin{tabular}{lcccc}
\br
 $Method$& $Q_{min}^2,\,\text{GeV}^2$ & $\mu_4^{p-n}/M^2$ &
 $\mu_6^{p-n}/M^4$ & $\mu_8^{p-n}/M^6$  \\
 \br
                     &  0.47 & $-0.055(3)$  &     0        &    0    \\
   NNLO APT          &  0.17 & $-0.062(4)$  &  0.008(2)    &    0       \\
   no evolution      &  0.10 & $-0.068(4)$  &  0.010(3)    & $-0.0007(3)$  \\
 \mr
                     &  0.47 & $-0.051(3)$  &     0        &    0        \\
   NNLO APT          &  0.17 & $-0.056(4)$  &  0.0087(4)   &    0        \\
   with evolution    &  0.10 & $-0.058(4)$  &  0.0114(6)   & $-0.0005(8)$  \\
\br
\end{tabular}
\end{center}
\end{table}
We do not take into account the RG-evolution in $\mu_4^{p-n}$ for the
standard PT calculations and compare with FAPT since the only
effect of that would be the enhancement of the Landau
singularities by extra divergencies at $Q^2\sim\Lambda^2$, whereas at
higher $Q^2\sim 1\,\text{GeV}^2$ the evolution is negligible with
respect to other uncertainties. We see from Table~\ref{tab:BjEv} that the
fit results become more stable with respect to $Q_{min}$
variations, which reduces the theoretical uncertainty of the BSR
analysis.

\section{Summary}
To summarize, APT and FAPT are the closed theoreti\-cal schemes
without unphysical singularities and additional phenomenological
pa\-ra\-me\-ters which allow one to combine RG-invariance,
$Q^2$-analyticity, compatibility with linear integral
trans\-for\-mations and essentially incorporate
nonper\-tur\-ba\-ti\-ve structures.
The APT provides a natural way for the coupling constant and related
quantities. These properties of the coupling constant are the universal loop-independent infrared limit and weak
dependence on the number of loops. At the same time, FAPT provides an
effective tool to apply the Analytic approach for RG improved
perturbative amplitudes. This approaches are used in many
applications. In particular, in this paper we consider the
application of APT and FAPT to the RG-evolution of nonsinglet
struc\-tu\-re functions and Bjorken sum rule higher-twist
ana\-ly\-sis at the scale $Q^2\sim\Lambda^2$ considered.

The sin\-gu\-la\-rity-free, finite couplings ${\cal
A}_{\nu}(Q^2),{\mathfrak A}_{\nu}(s)$ appear in APT/FAPT as
analytic images of the standard QCD coupling powers $\alpha_{\rm
s}^{\nu}(Q^2)$ in the Euclidean and Minkowski domains, respectively.
In this paper, we presented the theoretical background, used in a
package {\sf ``FAPT''}~\cite{KB13} based on the system {\sf
Mathematica} for QCD calcula\-tions in the framework of APT/FAPT,
which are needed to compute these couplings up to N$^3$LO of the RG
running. We hope that this will expand the use of these
approaches. \vspace*{1.5mm}

\textbf{Acknowledgments}\vspace*{+0.5mm}

I am grateful to the organizers and conveniers of the ACAT2013
workshop for the invitation and for arranging such a nice even. In
addition, I would like to thank \boxed{\text{Alexander
Bakulev}}\,, Sergei Mikhailov and Andrei Kataev for stimulating
discussions and useful remarks. This work was supported in part by
the Belarussian state fundamental research programm
``Convergency'', BelRFBR grants under Grant No.\ F12D-002 and No.\
F13M-143 and by RFBR Grant No.\ 11-01-00182.

\end{document}